\documentclass[fleqn,twoside]{article}
\usepackage{espcrc2}

\usepackage{graphicx}
\usepackage[figuresright]{rotating}

\newcommand{\AmS}{{\protect\the\textfont2
  A\kern-.1667em\lower.5ex\hbox{M}\kern-.125emS}}

\title{Lattice-QCD-based Schwinger-Dyson Approach for Chiral Symmetry Restoration 
at Finite Temperature}

\author{Hideaki~Iida\address[TITech]{Dept. of Physics, Tokyo Institute of Technology, Ohokayama 2-12-1, Meguro, Tokyo 152-8551, Japan},
Makoto~Oka\addressmark[TITech] and Hideo~Suganuma\addressmark[TITech]}

\begin{document}

\begin{abstract}
We propose the Schwinger-Dyson (SD) formalism based on lattice QCD, i.e., LQCD-based SD formalism, 
for the study of dynamical chiral-symmetry breaking in QCD.
We extract the kernel function $K(p^2)$ in the SD equation from the lattice data 
of the quark propagator in the Landau gauge. 
As remarkable features, we find infrared vanishing and intermediate enhancement 
of the kernel function $K(p^2)$ in the SD equation. 
We apply the LQCD-based SD equation to thermal QCD, 
and calculate the quark mass function at the finite temperature.
We find chiral symmetry restoration at the critical temperature $T_c \sim 100{\rm MeV}$.
\vspace{1pc}
\end{abstract}

\maketitle

\section{Introduction}

Dynamical chiral-symmetry breaking (DCSB) is one of the most important nonperturbative features in QCD.
For the study of DCSB, the Schwinger-Dyson (SD) formalism~\cite{H84,M93,SST95,BPRT03} 
has been used as an interesting and powerful method. 
For QCD, the SD formalism consists of an infinite series of nonlinear integral equations 
which determine the $n$-point Green's function of quarks and gluons,
and therefore it includes the infinite-order effect of the QCD gauge coupling constant $g$.

For instance, the SD equation for the quark propagator $S(p)$ is described with 
the nonperturbative gluon propagator $D_{\mu\nu}(p)$ and 
the nonperturbative quark-gluon vertex $g \Gamma_\nu(p,q)$ as 
\begin{eqnarray}
S^{-1}(p)=S^{-1}_0 
+g^2 \int_q \gamma_\mu S(q) D_{\mu\nu}(p-q) \Gamma_\nu(p,q),
\label{eqn:SDE1}
\end{eqnarray}
where $S_0(p)$ denotes the bare quark propagator and 
the simple notation $\int_q \equiv \int\frac{d^4 q}{(2\pi)^4}$ has been used in the Euclidean metric.

In the practical calculation for QCD, however, the SD formalism is drastically truncated: 
the perturbative gluon propagator and the one-loop running coupling are used instead of 
the nonperturbative quantities in the original formalism.
This simplification seems rather dangerous because some of 
nonperturbative-QCD effects are neglected.

In this paper, we formulate the SD equation based on the recent lattice QCD (LQCD) results, 
i.e., the LQCD-based SD equation,  and aim to construct 
a useful and reliable analytic framework including the proper nonperturbative effect in QCD.
Using the LQCD-based SD equation, we investigate also DCSB at finite temperatures.

\section{The Quark Propagator in Lattice QCD}

First, we briefly review the quark propagator $S(p)$ in lattice QCD.
The inverse quark propagator in the Landau gauge is generally given by  
\begin{eqnarray}
S^{-1}(p)=Z(p^2)/\{\not p+M(p^2)\}
\label{eqn:quarkprop}
\end{eqnarray}
in the Euclidean metric.
Here, $M(p^2)$ is called as the quark mass function, and 
$Z(p^2)$ corresponds to the wave-function renormalization of the quark field. 
In the quark propagator, DCSB is characterized by 
the mass generation as $M(p^2) \neq 0$.

The quark mass function $M(p^2)$ in the Landau gauge 
is recently measured in lattice QCD at the quenched level~\cite{BHW02}, 
and the lattice data in the chiral limit is well reproduced by 
\begin{eqnarray}
M(p^2)={M_0}/\{1+(p/\bar p)^\gamma\}
\label{eqn:quarkmass}
\end{eqnarray}
with $M_0$=260 MeV, $\bar p$=870MeV and $\gamma$=3.04.
The infrared quark mass $M(0)=M_0 \simeq 260{\rm MeV}$ seems consistent with the 
constituent quark mass in the quark model. 
Using this lattice result of $M(p^2)$, the pion decay constant is calculated as $f_\pi \simeq$ 87 MeV 
with the Pagels-Stokar formula~\cite{PS79}, 
and the quark condensate is obtained as $\langle \bar qq \rangle_{\Lambda =1{\rm GeV}} \simeq -(220{\rm MeV})^3$.
These quantities related to DCSB seem consistent with the standard values.

\section{The Schwinger-Dyson Equation}

In this section, we formulate the SD equation for quarks in the chiral limit in the Landau gauge.
By taking the trace Eq.(\ref{eqn:SDE1}), one finds 
\begin{eqnarray}
\frac{M(p^2)}{Z(p^2)}=\frac{g^2}{4}\int_q {\rm tr}\{\gamma_\mu \frac{Z(q^2)}{\not q+M(q^2)}\Gamma_\nu(p,q)\} D_{\mu\nu}(\tilde q)
\label{eqn:SDE2}
\end{eqnarray}
with $\tilde q \equiv p-q$.
For the quark-gluon vertex, 
we assume the chiral-preserving vector-type vertex, 
\begin{eqnarray}
\Gamma_\mu(p,q)=\gamma_\mu\Gamma((p-q)^2),
\label{eqn:vertex}
\end{eqnarray}  
which keeps the chiral symmetry properly.
(In contrast, to be strict, 
the Higashijima-Miransky approximation~\cite{H84,M93} explicitly breaks the chiral symmetry in the formalism.) 
Then, one obtains 
\begin{eqnarray}
\frac{M(p^2)}{Z(p^2)}=C_Fg^2\int_q  \frac{Z(q^2)M(q^2)}{q^2+M^2(q^2)}\Gamma(\tilde q^2) D_{\mu\mu}(\tilde q) 
\label{eqn:SDE3}
\end{eqnarray}
with $C_F=4/3$ being the color factor for quarks.
In the Landau gauge, the Euclidean gluon propagator is generally expressed by 
\begin{eqnarray}
D_{\mu\nu}(p^2)= \frac{d(p^2)}{p^2} \left( \delta_{\mu \nu} -\frac{p_{\mu} p_{\nu}}{p^2} \right),
\label{eqn:gluonprop}
\end{eqnarray}
where we refer to $d(p^2)$ as the gluon polarization factor. 
Therefore, Eq.(\ref{eqn:SDE3}) is rewritten as 
\begin{eqnarray}
\frac{M(p^2)}{Z(p^2)}=3C_Fg^2\int_q  \frac{Z(q^2)M(q^2)}{q^2+M^2(q^2)}\Gamma(\tilde q^2) \frac{d(\tilde q^2)}{\tilde q^2}. 
\label{eqn:SDE4}
\end{eqnarray}
Here, we define the kernel function  
\begin{eqnarray}
K(p^2) \equiv g^2 \Gamma(p^2) d(p^2)
\label{eqn:kernel}
\end{eqnarray}
as the product of the quark-gluon vertex $\Gamma(p^2)$ and the gluon polarization factor $d(p^2)$.
Then, the SD equation is expressed as   
\begin{eqnarray}
\frac{M(p^2)}{Z(p^2)}=3C_F\int_q  \frac{Z(q^2)M(q^2)}{q^2+M^2(q^2)} \frac{K((p-q)^2)} {(p-q)^2}.
\label{eqn:SDE5}
\end{eqnarray}
In the Landau gauge, the quark wave-function renormalization is not so significant and seems to be 
approximated as $Z(p^2)=1$, which reduces the SD equation to 
\begin{eqnarray}
M(p^2)=3C_F\int_q  \frac{M(q^2)}{q^2+M^2(q^2)} \frac{K((p-q)^2)} {(p-q)^2}. 
\label{eqn:SDE6}
\end{eqnarray}

\section{Extraction of the Kernel Function in the SD Equation from Lattice QCD}

In this section, we extract the kernel function $K(p^2)\equiv g^2 \Gamma(p^2)d(p^2)$ in the SD equation (\ref{eqn:SDE6}) 
using the quark mass function $M(p^2)$ obtained in lattice QCD. 
By shifting the integral variable from $q$ to $\tilde q \equiv p-q$, 
we rewrite 
Eq.(\ref{eqn:SDE6}) as 
\begin{eqnarray}
M(p^2)=3C_F \int_{\tilde q} \frac{M((p-\tilde q)^2)}{(p-\tilde q)^2+M^2((p-\tilde q)^2)} \frac{K(\tilde q^2)} {\tilde q^2}. 
\label{eqn:SDE7}
\end{eqnarray}
Therefore, we obtain 
\begin{eqnarray}
M(p^2)=\frac{3C_F}{8\pi^3}\int_0^\infty d \tilde q^2 \Theta(p, \tilde q) K(\tilde q^2), 
\label{eqn:SDE8}
\end{eqnarray}
where $\Theta(p,q)$ is defined with  $M(p^2)$ as 
\begin{eqnarray}
\Theta(p,q) \equiv \int_0^\pi d\theta \sin^2 \theta \hspace{3.5cm}
\nonumber \\ 
\frac{M(p^2+q^2-2pq\cos\theta)}{p^2+q^2-2pq\cos\theta+M^2(p^2+q^2-2pq\cos\theta)}. 
\label{eqn:Theta}
\end{eqnarray}
Since the quark mass function $M(p^2)$ is given by Eq.(\ref{eqn:quarkmass}) 
in lattice QCD, we can calculate the kernel function $K(p^2)$ from Eq.(\ref{eqn:SDE8}).

As shown in Fig.1, we numerically obtained the kernel function $K(p^2)\equiv g^2 \Gamma(p^2)d(p^2)$ extracted from the lattice QCD result 
for the quark propagator in the Landau gauge.
\begin{figure}[t]
\centering
\rotatebox{-90}{\includegraphics[width=2in]{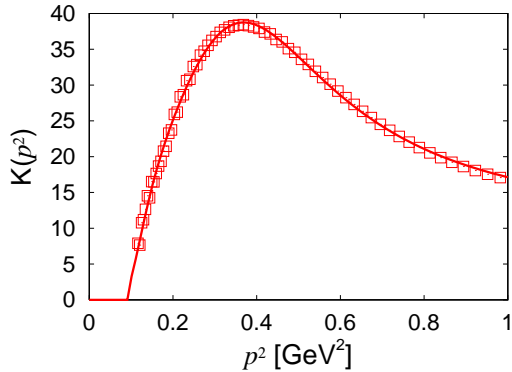}}
\vspace{-0.5cm}
\caption{
The kernel function in the SD equation, $K(p^2) \equiv g^2 \Gamma(p^2)d(p^2)$, 
extracted from the lattice QCD result of the quark propagator in the Landau gauge. 
The calculated data are denoted  by the square symbols, and  
the solid curve denotes a fit function for them.
In the very infrared region, the kernel seems consistent with zero.
As remarkable features, $K(p^2)$ exhibits infrared vanishing and intermediate enhancement.}
\vspace{-0.4cm}
\end{figure}

As remarkable features, 
we find ``infrared vanishing" and ``intermediate enhancement" 
in the kernel function $K(p^2)$ in the SD equation.
In fact, $K(p^2)$ seems consistent with zero in the very infrared region as 
\begin{eqnarray}
K(p^2 \sim 0)\simeq 0, 
\label{eqn:IRvanishing}
\end{eqnarray}
while $K(p^2)$ exhibits a large enhancement in the intermediate-energy region around $p \sim$ 0.5GeV.

These tendencies of infrared vanishing and intermediate enhancement in the kernel function 
$K(p^2)\equiv g^2 \Gamma(p^2)d(p^2)$  
are observed also in 
the direct lattice-QCD measurement for   
the polarization factor $d(p^2)$ in the gluon propagator 
in the Landau gauge~\cite{L98}.

Note here that the usage of the perturbative gluon propagator and the one-loop running coupling $g_{\rm run}(p^2)$
corresponds to $K(p^2)=g^2_{\rm run}(p^2)$, which largely differs from the present result based on lattice QCD 
both in the infrared and in the intermediate-energy regions.
In fact, the simple version of the SD equation using the perturbative gluon propagator and the one-loop running coupling 
would be too crude for the quantitative study of QCD.

\section{Chiral Symmetry at Finite Temperature}

Finally, we demonstrate a simple application of 
the LQCD-based SD equation to chiral symmetry restoration in finite-temperature QCD. 

At a finite temperature $T$, 
the field variables obey the (anti-)periodic boundary condition 
in the imaginary-time direction, which leads to 
the SD equation for the thermal quark mass $M_n({\bf p}^2)$ 
of the Matsubara frequency $\omega_n \equiv (2n+1)\pi T$ as 
\begin{eqnarray}
M_n({\bf p}^2)=\int_{m,{\bf q}} \frac{3C_F M_m({\bf q}^2)}{\omega_m^2+{\bf q}^2+M_m^2({\bf q}^2)} 
\frac{K(\omega_{nm}^2+\tilde {\bf q}^2)}
{\omega_{nm}^2+\tilde {\bf q}^2},
\label{eqn:SDET}
\end{eqnarray}
with
$\int_{m,{\bf q}} \equiv T \sum_{m=-\infty}^\infty \int \frac{d^3q}{(2\pi)^3}$, 
$\omega_{nm} \equiv \omega_n-\omega_m$ and 
$\tilde{\bf q} \equiv {\bf p}-{\bf q}$.

Using the kernel function $K(p^2)$ obtained in the previous section, 
we solve Eq.(\ref{eqn:SDET}) for the thermal quark mass $M_n({\bf p}^2)$.
Figure 2 shows 
the preliminary result for 
the thermal infrared quark mass $M_0({\bf p}^2=0)$ 
plotted against the temperature $T$.
We thus find chiral symmetry restoration at 
a critical temperature 
$T_c \sim 100{\rm MeV}$.

\vspace{-0.5cm}

\begin{figure}[hb]
\centering
\rotatebox{-90}{\includegraphics[width=2in]{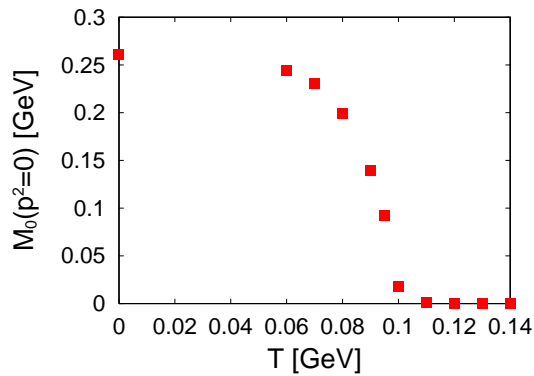}}
\vspace{-0.5cm}
\caption{
The thermal infrared quark mass $M_0({\bf p}^2=0)$ plotted against the temperature $T$.
The critical temperature $T_c$ is found to be about $100{\rm MeV}$.
}
\vspace{-0.5cm}
\end{figure}

\end{document}